\documentstyle[12pt] {article}

\newcommand{\beq}{\begin{equation}}
\newcommand{\eeq}{\end{equation}}
\newcommand{\beqa}{\begin{eqnarray}}
\newcommand{\eeqa}{\end{eqnarray}}

\begin{document}

\begin{center}
{\large \bf On the Convergence of the WKB Series \\
for the Angular Momentum Operator}
\vskip 1. truecm

{\bf Luca Salasnich}
\footnote{E--mail: salasnich@math.unipd.it}${}^{, \dag, \ddag, \S}$ and
{\bf Fabio Sattin}\footnote{
{\it Present address}: 
Istituto Gas Ionizzati del C.N.R.,
Corso Stati Uniti 4, 35127 Padova, Italy. E--mail: sattin@igi.pd.cnr.it}
${}^{, \P, \pounds}$ 
 
\vskip 0.3cm
${}^{\dag}$Dipartimento di Matematica Pura ed Applicata, Universit\`a di 
Padova, \\
Via Belzoni 7, 35131 Padova, Italy \\
${}^{\ddag}$Istituto Nazionale di Fisica Nucleare, Sezione di Padova, \\
Via Marzolo 8, 35131 Padova, Italy \\
${}^{\S}$Istituto Nazionale di Fisica della Materia, Unit\`a di Milano, \\
Via Celoria 16, 20133 Milano, Italy \\
${}^{\P}$Dipartimento di Ingegneria Elettrica, Universit\`a di Padova, \\
Via Gradenigo 6/a, 35131 Padova, Italy  \\
${}^{\pounds}$Istituto Nazionale di Fisica della Materia, Unit\`a di Padova, \\
Corso Stati Uniti 4, 35127 Padova, Italy
\end{center}
\vskip 0.5 truecm

{\bf Abstract.} In this paper we prove a recent conjecture 
[Robnik M and Salasnich L 1997 {\it J. Phys. A: Math. Gen.} {\bf 30} 1719] 
about the convergence of the WKB series for the angular momentum operator. 
We demonstrate that the WKB algorithm for the angular momentum 
gives the exact quantization formula if all orders are summed. 
Finally, we discuss the supersymmetric semiclassical quantum mechanics (SWKB), 
which gives the correct quantization of the angular momentum 
at the leading order. 

\vskip 0.2cm


\newpage

\section{Introduction}
\par
The semiclassical methods used to solve the Schr\"odinger problem are 
of extreme importance to understand the global behaviour of 
eigenfunctions and energy spectra, since they allow to obtain
analytic expressions. The leading semiclassical approximation 
(torus quantization) is just the first term of a certain 
$\hbar$--expansion, which is called WKB (Maslov and Fedoriuk 1981). 
\par
Recently it was observed (Prosen and Robnik 1993, Graffi, 
Manfredi and Salasnich 1994, Robnik and Salasnich 1997a--in the 
following this work will be referred to as I) that the 
torus quantization generally fails to predict
the individual energy levels (and the eigenstates) within a vanishing 
fraction of the mean--energy level spacing. This conclusion is believed 
to be correct for general systems, including the chaotic ones. Therefore, 
a systematic study of the accuracy of semiclassical approximation is very 
important, especially in the context of quantum chaos (Casati  and Chirikov 
1995, Gutzwiller 1990). Since this is a difficult task, it has been 
attempted for simple systems, where in a few cases 
even exact solutions may be worked out (Dunham 1932, Bender, Olaussen and Wang 
1977, Voros 1993, Robnik and Salasnich 1997a). 
\par 
Robnik and Salasnich (1997b) 
(this work will be referred to as II) 
dealt with the WKB expansion for the Kepler problem: it was 
proved that an exact result is obtained once all terms 
are summed. In particular, 
the torus quantization (the leading WKB term) 
of the full problem is exact, even if the individual torus 
quantization of the angular momentum and of the radial Kepler 
problem separately are not, because the quantum 
corrections (i.e. terms higher than the torus quantization)
compensate mutually term by term. 
In the paper II Robnik and Salasnich had to do a conjecture 
about the higher terms of the WKB expansion. This conjecture is perfectly 
reasonable but not rigorously proved. In this work our goal is to prove 
that the same result of II can be reached rigorously 
by means of a slightly modified procedure. 
\par
In the framework of the supersymmetric 
semiclassical quantization (SWKB), Comtet, Bandrauk and Campbell 
(1985) obtained at the leading order the exact quantization of the radial 
part of the Kepler problem by using the correct value 
$L^2 = \hbar^2 l (l + 1)$. In the last section we complete the 
result of Comtet, Bandrauk and Campbell (1985). In fact, we show that 
also the exact quantization of the angular momentum is obtained 
at the first order of the SWKB expansion. 

\section{Eigenvalue problem for the angular momentum}
\par
The eigenvalue equation of the angular momentum operator 
(Landau and Lifshitz 1977) is 
\beq
\hat{L}^2 Y(\theta , \phi) = \lambda^2 \hbar^2 Y(\theta , \phi) \; ,
\eeq
with
\beq
\hat{L}^2 = \hat{P}_\theta^2 + { \hat{P}_{\phi}^2 \over \sin^2(\theta)} 
          = - \hbar^2 \left( { \partial^2 \over \partial \theta^2 } +
\cot(\theta) {\partial \over \partial \theta} \right) - \hbar^2
{ 1\over \sin^2(\theta)}  {\partial^2 \over \partial \phi^2} \; . 
\eeq
After the substitution
\beq
Y(\theta , \phi ) = T(\theta) e^{ i m \phi} \; ,
\eeq
we obtain
\beq
T''(\theta) + \cot(\theta) T'(\theta) + \left( \lambda^2 - 
{ m^2 \over \sin^2(\theta) } \right) T(\theta) = 0 \; .
\eeq
We shall consider the azimuthal quantum number $m$ as fixed. As well known, 
Eq. (4) is exactly solvable. 
Its eigenvalues and eigenfunctions are known from 
any text of quantum mechanics (see, {\it e.g.} Landau and Lifshitz 1977):
the former are $\lambda^2=l(l+1)$, $l \geq m$; the latter are the 
associate Legendre polynomials. 
\par
The WKB expansion for Eq. (4) has been studied in II; 
it was shown that higher--order terms quickly 
increase in complexity. The method of solution is to find an analytical 
recursive expression for all the higher--order terms, to sum the 
entire infinite series, and show that it is convergent to the exact
result. Instead of the original 
function $T$ we shall use the associated function $F$:
\beq
T(\theta) = { F(\theta) \over \sqrt{ \sin(\theta)} } \; ,
\eeq
from which we obtain
\beq
F''(\theta) + \left[ \left( \lambda^2 + {1 \over 4} \right) +
{ 1 \over \sin^2(\theta) } \left( { 1\over 4 } - m^2 \right)
\right] F(\theta) = 0 \; .
\eeq
This equation has the standard form of the one--dimensional Schr\"odinger 
equation with $ \hbar = 2 M \equiv 1 $. Its eigenvalues are 
$ (\lambda^2 + 1/4)$. 
We make the substitution of variable: $x=\theta + \pi/2$,
and the positions $U = m^2 - 1/4 , E = \lambda^2 + 1/4 $. Then 
Eq. (6) becomes
\beq
- F''(x) + {U \over \cos^2(x)} F (x)= E F(x)  \; .
\eeq
This is the main result of our paper, because 
the problem of the WKB quantization of Eq. (7) has already been dealt with 
in I. 
As we shall show, from I one proves that: i) Eq. (7) 
can be solved exactly ; ii) a semiclassical expansion of (7)
may be carried on to all orders ({\it i.e.} all terms may be exactly 
and analytically computed and summed); iii) the exact and the semiclassical
eigenvalues are the same. 

\section{WKB series for the angular momentum}
\par
We observe that in Eq. (7) does not appear $\hbar$, therefore 
an expansion in powers of this parameter is not possible. To 
override this difficulty a small parameter $\epsilon$ is introduced: 
\beq
- \epsilon^2 F''(x) + {U \over \cos^2(x)} F (x)= E F(x)  \; .
\eeq
This parameter $\epsilon$, which will be set to 1 at the end of the 
calculation, has formally the same role of $\hbar$ as ordering parameter. 
It has already been used in II to deal with the WKB expansion of (4). 
The formal WKB expansion for $F$ reads: 
\beq
F(x) = \exp \left( {i \over \epsilon} {\sum_{n=0}^{\infty} 
\sigma_n(x) \epsilon^n} 
\right) \; ,
\eeq
and we obtain a recursion relation for the phases: 
\beq
(\sigma_0'(x))^2 =  E - { U \over  \cos^2(x) } \; ,
\eeq
\beq
{ \sum_{k = 0}^n \sigma_k'(x) \sigma_{n-k}'(x) } + \sigma_{ n - 1}''(x) 
= 0 \; , 
\;\;\;\; n > 0 \; .
\eeq
The quantization condition is obtained by requiring that the wavefunction 
be single valued:
\beq
\oint d \sigma = { \sum_{k = 0}^{\infty} \oint d \sigma_k } = 
2 \pi  n_{\theta}  \; ,
\eeq
where $n_{\theta}$ is an integer number. All odd terms higher than the first 
vanish when integrated along the closed contour since they are exact 
differentials (Bender, Olaussen and Wang 1977)
\beq
\oint d\sigma_{2 k + 1} = 0 \; , \;\;\;\; k > 0 \; .
\eeq
It may be proved by induction (see I) that the solution of (10--11) is 
\beq
\sigma_n'(x) = (\sigma_0')^{1 - 3 n} P_n(\cos( x)) \sin^{f(n)}(x) \; ,
\eeq
with $ f(n) = 0$ for $n$ even, $f(n) = 1$ for $n$ odd, 
\beq
P_n(\cos(x)) = {\sum_{l=0}^{g(n)} C_{n , l} \cos^{2 l - 3 n}(x) } \; ,
\eeq
with $ g(n) = (3 n - 2)/2$ for $n$ even, $ g(n) = (3 n - 3)/2$ for $n$ 
odd, $ C_{0 , 0} = 1 , C_{1 , 0} = U/2 ,  
C_{2 k , 0} = (-1)^k (U/2)^{2 k} \left({}^{1/2}_k \right)$, and 
$ C_{ 2 k + 1, 0} = 0 , k > 0$. It is not necessary to know the 
value of the other coefficients since one finds that 
all the terms proportional to $ C_{n , l}$, $l > 0$ 
disappear after integration. 
The integral (12) becomes (see I for more details)
\beqa
\oint d \sigma &=& \oint d \sigma_0 + \oint d \sigma_1 +
{\sum_{k > 0}^{\infty} \oint d \sigma_{2 k} }  \\
&=& 2 \pi \left( \sqrt{E} - \sqrt{U} \right) - \pi  -
{\sum_{k > 0}^{\infty} {1 \over 2} \left( {1 \over 2} \choose k \right) 
{ 2 \pi \over \left(\sqrt{ 4 U}\right)^{2 k -1} } }\\
&=& 2 \pi \left( \sqrt{E} - {1 \over 2} - 
{\sum_{k = 0}^{\infty} {1 \over 2} \left( {1 \over 2} \choose k \right) 
{ 2 \pi \over \left(\sqrt{ 4 U}\right)^{2 k -1} }} \right) = 
2 \pi n_{\theta} \; ,
\eeqa 
but $\sum_{k = 0}^{\infty} \left( {1 \over 2} \choose k \right) x^{1 - 2 k} =
\sqrt{1 + x^2}$ for $|x|>1$, therefore Eq. (18) reads 
\beq
\sqrt{E} - {1 \over 2} \sqrt{1 + 4 U^2} - { 1 \over 2} =
\sqrt{\lambda^2 + {1\over 4}} - { 1 \over 2} \sqrt{ 4 m^2} - {1 \over 2} =
n_{\theta} \; .
\eeq
Now, because $E=\lambda^2+1/4$ and $U=m^2-1/4$, we obtain 
\beq
\lambda^2 = \left( m + n_{\theta} + {1 \over 2} \right)^2 - { 1 \over 4} 
=( m + n_{\theta} ) ( m + n_{\theta} + 1) \; ,
\eeq
and, with the position $l=n_{\theta}+m$, we have
\beq
\lambda^2 = l (l+1) \; ,
\eeq
which is the expected result. Please note that the WKB series 
is convergent for $|x|>1$, thus for $m>0$. 
\par
We observe that the $\epsilon$--expansion is equivalent to the 
$1/U$--expansion (this is clear from the structure of Eq. 8). 
In the limit $U\to \infty$ it is easy to get the WKB expansion 
to the first order, which gives $\lambda^2 = (l+1/2)^2$, 
{\it i.e.} the torus quantization 
of the angular momentum (Langer 1937). 

\section{SWKB quantization of the angular momentum}
\par
To perform the supersymmetric semiclassical quantization 
(SWKB) of Eq. (4) or (6), it is necessary to know the ground state 
wave--function $T_0(\theta )=\sin^m{(\theta )}$ 
and its eigenvalue $\lambda_0=m(m+1)$. 
Then we can define the supersymmetric (SUSY) potential
\beq
\Phi(\theta) = - { d \ln ( F_0 (\theta)) \over d \theta} =
 - \left(  m + { 1 \over 2} \right) \cot(\theta) \; ,
\eeq
with
\beq
F_0(\theta) = T_0(\theta) \sqrt{\sin(\theta)} \; .
\eeq 
From $\Phi$ the two SUSY partner potentials and Hamiltonians
may be defined
\beq
H_{\pm} = - {d^2 \over d \theta^2} + V_{\pm}(\theta) \; ,
\eeq
\beq
V_{\pm}(\theta) = \Phi^2(\theta)  \pm \Phi'(\theta) \; .
\eeq
It is possible to prove (see Junker 1996 and references therein for details) 
that: i) the ground-state energy of $H_{-}$,
$E_{-}^0$, vanishes; ii) all other eigenvalues of $H_{-}$, $E_{-}$, coincide 
with that of $H_{+}$; iii) the spectrum of $H_{-}$ and that of (7)
differ by a constant:
\beq
E_{-} = \lambda^2 + { 1\over 4} - \left( \lambda_0^2 + { 1\over 4}
\right) \; , 
\eeq
where $\lambda_0=m(m+1)$ is the eigenvalue 
of the ground state of Eq. (4). 
\par 
Now we apply the SWKB formalism to $H_{-}$ of Eq. (24). 
At the leading order one gets
\beq
{\int_a^b \sqrt{ E_{-} - \Phi^2(x)} \; d x } = n_{\theta} \pi
\eeq
with $a , b$ roots of 
\beq
E_{-} - \Phi^2(x) = 0
\eeq
This formula is also referred to as CBC formula, from
Comtet, Bandrauk and Campbell (1985). We observe that on the left hand
side of the previous formulas $\Phi^2$ appears instead of the full 
potential $V_{-}$. 
\par
From Eqns. (22), (27) and (28) one easily finds
\beq
 \sqrt{ E_{-} + \left( m + {1 \over 2} \right)^2 }
- \left( m + {1 \over 2} \right)  \\
= n_{\theta} \; ,
\eeq
with $b = - a = \arctan \sqrt{ \left( m + { 1 \over 2} \right) 
\over E_{-} }$. By inverting the previous formula we have
\beq
E_{-} = \left( n_{\theta} + m + {1 \over 2} \right)^2 -
\left( m + {1 \over 2} \right)^2 \; , 
\eeq
and, by using Eq. (26) with $\lambda_0 =m (m+1)$, we get  
\beq
\lambda^2 = (n_{\theta} + m) ( n_{\theta} + m + 1) \; ,
\eeq
which yields the exact quantization, after the position $l = n_{\theta} + m$.

\section{Conclusions}
\par
The three--dimensional central potentials are fundamental 
in physics, and also the semiclassical
treatment of them has implications in many fields: factorization properties of 
the one--dimensional potentials (Infeld and Hull 1957), 
general properties of the semiclassical quantization 
of the systems with more than one degree
of freedom, both integrable or not. Nevertheless, until the
paper of Robnik and Salasnich (1997b), no detailed study had been done 
on half of the problem, the WKB quantization of the angular part. 
Our present paper completes that work because 
it gives a rigorous proof of the convergence of the WKB series 
to the exact result. Moreover, in the last section, we have 
demonstrated that, by using SUSY quantum
mechanics, the eigenvalue problem of the angular momentum operator can 
be solved exactly at the lowest order within the semiclassical approximation.

\vskip 0.5 truecm

LS thanks Marko Robnik for many enlightening discussions. 
FS has been supported during this work by a grant of the Italian MURST. 

\newpage

\section*{References}
\parindent=0. pt

Barclay D T and Maxwell C J 1991 {\it Phys. Lett.} A {\bf 157} 357

Bender C M, Olaussen K and Wang P S 1977 {\it Phys. Rev.} D {\bf 16} 
   1740

Casati G and Chirikov B V 1995 {\it Quantum Chaos} (Cambridge: Cambridge
  University Press)

Comtet A, Bandrauk A D, and Campbell D K 1985 {\it Phys. Lett.} B
{\bf 150} 159

Dutt R, Khare A, and Sukhatme U P 1986 {\it Phys. Lett.} B {\bf 181} 295

Dunham J L 1932 {\it Phys. Rev.} {\bf 41} 713

Graffi S, Manfredi V R, Salasnich L 1994 {\it Nuovo Cimento} B 
   {\bf 109} 1147 

Gutzwiller M C 1990 {\it Chaos in Classical and Quantum Mechanics} (New York:
  Springer)

Infeld L and Hull T H 1957 {\it Rev. Mod. Phys.} {\bf 23} 21
 
Junker G 1996 {\it Supersymmetric Methods in Quantum and Statistical
 Physics} (Springer)

Landau L D and Lifshitz E M 1977 {\it Quantum Mechanics} (3rd Edition,
Pergamon)

Langer R I 1937 {\it Phys. Rev.} {\bf 51} 669

Maslov V P and Fedoriuk M V 1981 {\it Semi-Classical Approximations in 
Quantum Mechanics} (Reidel Publishing Company)

Prosen T and Robnik M 1993 {\it J. Phys. A: Math. Gen.} {\bf 26} L37

Robnik M and Salasnich L 1997a {\it J. Phys. A: Math. Gen.} {\bf 30}
   1711 

Robnik M and Salasnich L 1997b {\it J. Phys. A: Math. Gen.} {\bf 30}
   1719 

\end{document}